\documentclass[12pt,preprint]{aastex}

\shorttitle{Elliptical Galaxies}
\shortauthors{Burkert, Naab  \& Johansson}

\begin{document}

\title{SAURON's Challenge for the Major Merger Scenario of Elliptical Galaxy Formation}

\author{Andreas Burkert\altaffilmark{1}, Thorsten Naab\altaffilmark{1} \& Peter H. Johansson\altaffilmark{1}}

\altaffiltext{1}{University Observatory Munich, Scheinerstrasse 1, D-81679 Munich, 
Germany}

\email{burkert@usm.uni-muenchen.de, naab@usm.uni-muenchen.de, pjohan@usm.uni-muenchen.de}

\newcommand\msun{\rm M_{\odot}}
\newcommand\lsun{\rm L_{\odot}}
\newcommand\msunyr{\rm M_{\odot}\,yr^{-1}}
\newcommand\be{\begin{equation}}
\newcommand\en{\end{equation}}
\newcommand\cm{\rm cm}
\newcommand\kms{\rm{\, km \, s^{-1}}}
\newcommand\K{\rm K}
\newcommand\etal{{\rm et al}.\ }
\newcommand\sd{\partial}

\begin{abstract}

The intrinsic anisotropy $\delta$ and flattening $\epsilon$ of simulated merger remnants
is compared with elliptical galaxies that have been observed by the SAURON collaboration,
and that were analysed using axisymmetric Schwarzschild models. Collisionless binary mergers of stellar
disks and disk mergers with an additional isothermal gas component, neglecting star formation
cannot reproduce the observed trend $\delta = 0.55 \epsilon$ (SAURON relationship).
An excellent fit of the
SAURON relationship for flattened ellipticals with $\epsilon \geq 0.25$ is however found
for merger simulations of disks with gas fractions $\geq 20\% $, including star formation
and stellar energy feedback. Massive black hole feedback does not strongly affect this result.
Subsequent dry merging of merger remnants however does not generate
the slowly-rotating SAURON ellipticals which are characterized by low ellipticities
$\epsilon  < 0.25$ and low anisotropies. This indicates that at least some
ellipticals on the red galaxy sequence did not form by binary mergers of disks or
early-type galaxies. We show that stellar spheroids resulting from multiple,
hierarchical mergers of star-bursting subunits in a cosmological context
are in excellent agreement with the low ellipticities and anisotropies
of the slowly rotating SAURON ellipticals and their observed trend of $\delta$ with $\epsilon$.
The numerical simulations indicate that the SAURON relation might be a result of strong
violent relaxation and phase mixing of multiple, kinematically cold stellar subunits
with the angular momentum of the system determining its location on the relation.
\end{abstract}


\section{Introduction}
A popular formation scenario for early-type
galaxies is the collision and merger of two roughly
equal-mass galaxies with mass ratios between 1:1 and 4:1. This
famous major merger scenario \citep{1972ApJ...178..623T} has
been very successful in explaining observed properties of
ellipticals, like their kinematics, surface density profile or
isophotal shape 
\citep{1981MNRAS.197..179G,1983MNRAS.205.1009N,1988ApJ...331..699B,
1990Natur.344..379B,1992ARA&A..30..705B,1992ApJ...400..460H,
1993ApJ...409..548H,1993A&A...278...23B,1999ApJ...523L.133N,2001ApJ...554..291C, 
2003ApJ...597..893N,2005MNRAS.357..753G,2005MNRAS.360.1185J,
2005A&A...437...69B,2006MNRAS.369..625N,2006MNRAS.372..839N,
2006ApJ...641...21R,2006astro.ph..7446C}. 
Numerical simulations for example showed that
the family of disky, fast rotating ellipticals could
result from stellar disk galaxy mergers with unequal mass ratios of 3:1
to 4:1 \citep{1998giis.conf..275B,1998ApJ...502L.133B,1999ApJ...523L.133N,
2003ApJ...597..893N,2005A&A...437...69B,2006MNRAS.372..839N} or from
gas-rich 1:1 to 2:1 disk mergers where the gas 
subsequently settles into the equatorial plane of the merger remnant and
produces a new stellar disk component \citep{2005MNRAS.359.1379K,
1996ApJ...471..115B,2001ApJ...555L..91N,2002MNRAS.333..481B,
2005ApJ...622L...9S}. 
Boxy, slowly rotating ellipticals, on the other hand form in
stellar disk-disk mergers with mass ratios of 1:1 to 2:1
\citep{1994ApJ...427..165H,2003ApJ...597..893N} or from multiple major
disk mergers \citep{1996ApJ...460..101W}. \citet{2007arXiv0709.3439B} showed that
repeated minor mergers result in remnant properties very similar to one
corresponding to major mergers.  

A serious problem of the major merger scenario was the fact
that cosmological models do not predict a dependence of the mass ratio
of mergers on total galaxy mass or luminosity \citep{2003ApJ...597L.117K}.
If, on the other hand, the mass ratio determines the isophotal shape and
rotational properties of merger remnants one would expect that the ratio
of the number of fast rotating, disky ellipticals to the number of slowly
rotating, boxy systems should be independent of luminosity
\citep{2006ApJ...636L..81N}. This is in contrast with observations that show 
a strong dependence of isophotal shape and rotational properties
on galaxy mass. While massive galaxies are preferentially boxy, slow rotators,
lower-mass ellipticals are predominantly disky and fast rotators
(for a summary see \citealp{1996ApJ...464L.119K}).
\citet{2003ApJ...597L.117K} argued that this mass dependence of galaxy properties
could be explained as a result of differences in the morphologies of
the merging progenitors (see e.g. \citealp{2007MNRAS.tmp..808K}). Using semi-analytical models they
showed that  
gas-rich disk-disk mergers dominate at the low-mass end of ellipticals while
intermediate mass ellipticals should have formed preferentially from mixed 
mergers involving a disk and an elliptical galaxy. Finally, the most massive
early-type galaxies should have experienced a last elliptical-elliptical merger (dry merger)
\citep{2006ApJ...636L..81N}.
Mixed mergers have not yet been studied in details, although, according to
\citet{2003ApJ...597L.117K} they should be more frequent than dry mergers
(see however \citealp{2007arXiv0706.1243H}). Dry mergers
and their implications for the formation of the red galaxy sequence
have however received a lot of attention recently
\citep{2005MNRAS.359.1379K,2005MNRAS.361.1043G,2005astro.ph..6044F,
2006ApJ...636L..81N,2006MNRAS.369.1081B}. 

Further refinement of theoretical models has recently been achieved by including
black-hole physics in simulations of galaxy mergers and elliptical galaxy
evolution. Energetic feedback from central black holes might
solve some pending problems of major merger
models, like the suppression of late inflow of cold gas and star formation
that would make ellipticals look much
bluer than observed \citep{2005ApJ...620L..79S}. 

In summary, despite several still unsolved questions (e.g. \citealp{2007astro.ph..2535N}),
the major merger scenario has  become a popular model in order to explain the origin of
bulge-dominated, spheroidal galaxies 
(see e.g. \citealp{2007arXiv0706.1246H,2007arXiv0706.1243H}).

Progress in understanding galactic evolution is often driven by 
strong interactions between observers and theorists.
Increasingly more sophisticated theoretical/numerical models are confronted 
with continuously improving observations that
lead to new theoretical challenges. One example is the
SAURON project \citep{2004MNRAS.352..721E} which aims to determine the 
2-dimensional structural and kinematical properties of 
early-type galaxies using a panoramic integral-field
spectrograph. In order to interpret the observations
and study the intrinsic galaxy structure,
axisymmetric Schwarzschild models are applied to the
observations \citep{2006MNRAS.366.1126C,2007MNRAS.379..418C}. The results,
published so far, have revealed interesting fine structures and physical properties which
provide new and deeper insight into the origin of galaxies, in
particular when compared to simulation
\citep{2000MNRAS.316..315B,2007MNRAS.376..997J}.

In this paper we confront a recently published SAURON analysis of preferentially
axisymmetric elliptical galaxies \citep{2007MNRAS.379..418C} with the
predictions of numerical merger simulations and cosmological models of
galaxy formation. Section 2 summarizes the observations. Section 3
shows that simulations of collisionless and gaseous disk-disk mergers,
neglecting star formation and stellar feedback cannot reproduce the observational results.
We demonstrate that star formation and stellar energy feedback has a strong effect
on the final structure of merger remnants, leading to a good agreement with the SAURON
observations of fast rotating ellipticals. The origin of the round, almost isotropic,
slowly rotating SAURON ellipticals is explored in section 4. Isolated, dry mergers
of ellipticals that formed as discussed in section 3 cannot explain these objects.
We show however that cosmological initial conditions, leading to a series of multiple major and minor
mergers, coupled with local star bursts generate spheroidal stellar systems in very good
agreement with the observations. Conclusions follow in section 5.

\section{SAURON results}
\citet{2007MNRAS.379..418C} analyzed a subsample of 24 elliptical and 
lenticular galaxies of the SAURON survey which were biased against
triaxiality and consistent with being axisymmetric stellar systems.
Axisymmetric three-integral Schwarzschild models
(Schwarzschild 1979; Cappellari et al. 2005) were used to determine
the distribution of stellar orbits. For an investigation of the validity and accuracy
of axisymmetric Schwarzschild models in reproducing the intrinsic properties of
simulated merger remnants, see \citet{2007arXiv0708.2205T}. 
As SAURON integral-field kinematics was only
taken within the effective radius, the Schwarzschild calculations were restricted to a determination
of the orbital parameters for the 50\% most bound stars in each galaxy. From this,
global galactic parameters were derived.
The authors found two classes of spheroid-dominated galaxies, one with and 
one without significant
amount of specific angular momentum (Emsellem et al. 2007). They called
these groups slow and fast rotators, respectively. 
Following Binney (1978), for each galaxy the anisotropy parameter

\begin{equation}
\delta \equiv \frac{\Pi_{xx}-\Pi_{zz}}{\Pi_{xx}} .
\end{equation}

\noindent was determined. Here, the z axis coincides with the symmetry 
and rotation axis of the axisymmetric galaxy,

\begin{equation}
\Pi_{ii}=\int \rho \sigma_i^2 d^3x
\end{equation}

\noindent is the diagonal element of the velocity dispersion tensor in the i-th direction
\citep{1987gady.book.....B},
$\sigma^2_i=\langle v_i^2 \rangle - \langle v_i \rangle^2$ is the local mean velocity
dispersion in the i-th direction and $\rho$ is the local stellar density. 
For axisymmetric systems $\Pi_{xx} = \Pi_{yy}$. In addition, the so called intrinsic ellipticity 
$\epsilon_{int}$ was determined as the edge-on ellipticity of the projected
early-type galaxy, corrected for inclination effects. 

\begin{figure}[!ht]
\plotone{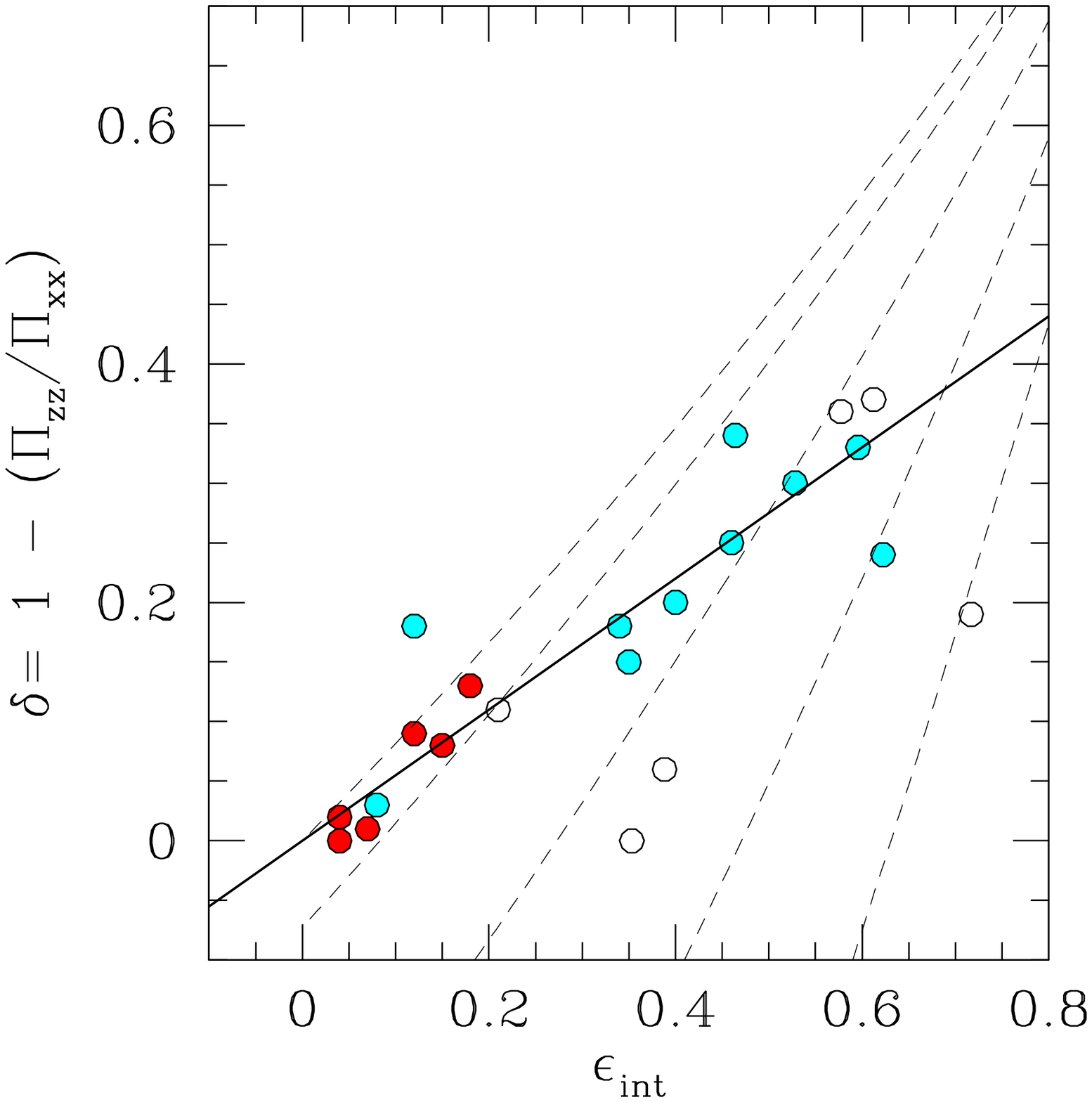}
\caption{
 \label{fig}
The anisotropy $\delta$ versus the edge-on projected ellipticity
$\epsilon_{int}$ is shown for SAURON ellipticals. The effect
of rotation on the ellipticity is demonstrated by the
dashed lines which from left to right show the theoretically predicted 
correlation $\delta (\epsilon)$ (Eqs. 4 and 5) for constant values of V$/\sigma$=0,0.25,0.5,0.75,1,
respectively. Red and cyan filled circles show slowly and fast rotating SAURON ellipticals, respectively.
Open circles show objects, classified as S0 galaxies.}
\end{figure}

The red and cyan filled circles in Fig. 1 show $\delta$ versus $\epsilon_{int}$
of the slow and fast rotating ellipticals of the SAURON sample, respectively.
The population of slow rotators
is intrinsically quite round ($\epsilon_{int} \leq 0.25$)
and characterized by an almost isotropic velocity dispersion with $\delta \leq 0.15$.
In contrast, the fast rotators are in general much flatter ($\epsilon_{int} \geq 0.3$) and more
anisotropic with $\delta \geq 0.15$. This result is in conflict with the
standard paradigm that fast rotating ellipticals are
nearly isotropic systems while slowly rotating ellipticals are
strongly anisotropic \citep{1978MNRAS.183..501B}.
\citet{2005MNRAS.363..597B} showed that projection
effects play an important role for fast rotators and that even strongly
anisotropic stellar systems would appear isotropic when viewed under random projection.
The origin of the isotropic, slowly rotating population is however puzzling. 
Despite the fact that the analyzed SAURON sample is still small, Fig. 1 clearly reveals
a strong correlation between $\delta$ and 
$\epsilon_{int}$ which can be fitted by the empirical relationship (solid line)

\begin{equation}
\delta = (0.55 \pm 0.1) \epsilon_{int}.
\end{equation}

Ellipticals are usually analysed, using the classical
(V/$\sigma$)-$\epsilon$ anisotropy diagram \citep{1978MNRAS.183..501B,2005MNRAS.363..937B} 
with V and $\sigma$ being the maximum projected rotational velocity and
projected central velocity dispersion, respectively, and $\epsilon$ the
apparent projected ellipticity. \citet{2005MNRAS.363..937B}
demonstrated that for axisymmetric systems, seen edge-on,  
the anisotropy is given by 

\begin{equation}
\delta = 1 - \frac{1+(V/\sigma)^2}{q(e)[1-\alpha (V/\sigma)^2]}
\end{equation}

\noindent where $\alpha$ measures the shear in the stellar streaming
velocity and 

\begin{equation}
q(e) \equiv \frac{0.5}{1-e^2} \times \frac{\arcsin e - e \sqrt{1-e^2}}
{\frac{e}{\sqrt{1-e^2}}- \arcsin e}
\end{equation}

\noindent with $e = (1-(1- \epsilon )^2)^{1/2}$.
$\alpha$ is in general small. The dashed curves in Fig. 1 show
the expected correlation between $\delta$ and $\epsilon$ for different
values of V/$\sigma$, adopting $\alpha = 0.15$ (Cappellari et al. 07). 
Due to the effect of rotational flattening, $\epsilon$ increases 
with increasing V/$\sigma$ for a given value of $\delta$. The fast rotating
SAURON ellipticals (cyan circles) cluster around V/$\sigma \approx 0.5$ while
the slowly rotating sample (red circles) is characterized by V/$\sigma \leq 0.25$.

Open circles in Fig. 1 correspond to objects classified as S0 galaxies.
Two of these objects have properties that
are similar to fast rotating ellipticals, indicating either a similar
origin or classification problems. Note however the 3 highly elliptical and fast rotating outliers 
that indicate that S0s are at least sometimes more rotationally dominated than 
fast rotating early-type galaxies.

In the next sections we will explore the question whether these observational
results are in agreement with the major merger scenario of early-type galaxy formation.

\section{The Origin of the Fast Rotating SAURON Ellipticals}

We start with an analysis of a
large sample of collisionless merger remnants of disk galaxies with different mass ratios
and initial orientations.
The progenitors consisted of a stellar disk, a stellar bulge
and a surrounding dark matter halo. Gas and star formation has been neglected.
Details of the initial conditions,
the simulations and the properties of the merger remnants have been presented elsewhere
\citep{2003LNP...626..327B,2003ApJ...597..893N,2006A&A...445..403K,2006MNRAS.372..839N}. There
it was shown that equal-mass mergers
with progenitor mass ratios of 1:1 to 2:1 produce slowly rotating and often boxy
remnants, resembling massive ellipticals, while unequal-mass mergers with
mass ratios of 3:1 - 4:1 generate fast rotating, disky remnants, resembling lower-mass
ellipticals. 

\begin{figure}[!ht]
\plotone{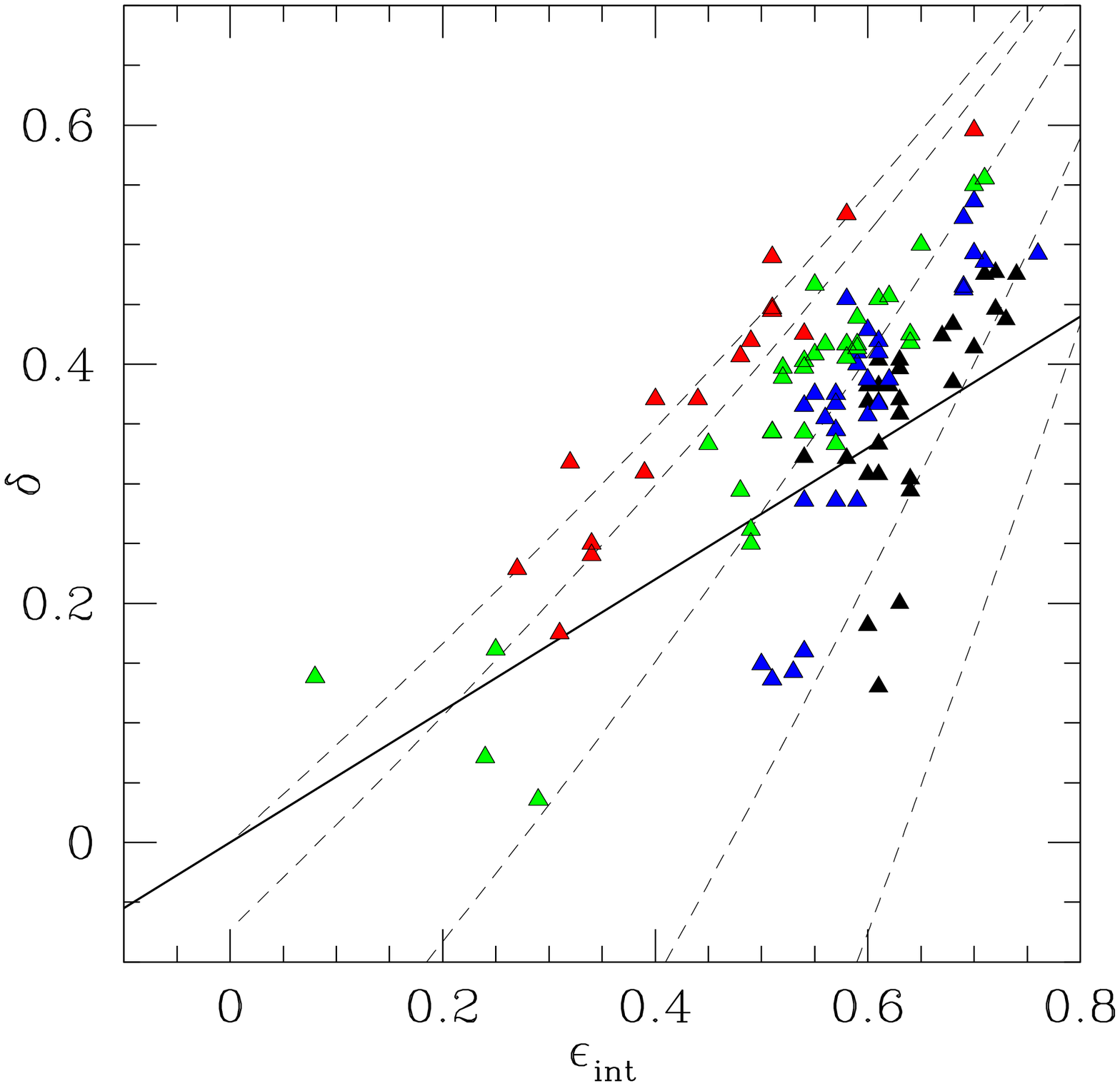}
\caption{
 \label{fig}
The structure of dissipationless merger
remnants (triangles) is compared with the SAURON relation, represented by
the solid line (dashed lines, see Fig.~1).
The red, green, blue and black filled triangles correspond to dynamically relaxed
merger remnants of collisionless
stellar disk mergers with mass ratios 1:1, 2:1, 3:1 and 4:1, respectively.}
\end{figure}

In order to match the SAURON analysis,
$\delta$ was determined from the 50\% most bound stellar particles 
of the relaxed merger remnants and $\epsilon_{int}$ from the edge-on projected 
stellar distribution. \citet{2007MNRAS.379..418C} used axisymmetric
models where $\Pi_{xx}=\Pi_{yy}$. One should however note that especially 1:1 mergers
are quite triaxial. Our 1:1 mergers have a roughly homogeneous distribution 
of the ratio $\Pi_{yy}/\Pi_{xx}$ which lies in the range of
$\Pi_{yy}/\Pi_{xx} = 0.75 - 1$. We therefore
use an averaged value of the velocity dispersion in the 
equatorial plane in order to determine the anisotropy:

\begin{equation}
\delta \equiv \frac{0.5(\Pi_{xx}+\Pi_{yy})-\Pi_{zz}}{0.5(\Pi_{xx}+\Pi_{yy})}
\end{equation}

\noindent which in the axisymmetric case reduces to Eq. 1.

Fig. 2 shows the distribution of all collisionless merger remnants in the
$\delta$-$\epsilon_{int}$-diagram. In agreement with previous work,
1:1 remnants (red triangles) are slowly rotating (V/$\sigma \leq 0.25$), consistent
with the slow rotators found by the SAURON team. 2:1 - 4:1 merger remnants
fall into the regime of $0.25 \leq$ V/$\sigma \leq 0.75$, consistent
with fast rotating SAURON ellipticals. 
However, in contrast to the SAURON observations,
no correlation of $\delta$ with $\epsilon$ is visible. Independent of
V/$\sigma$, $\epsilon$ or the progenitor mass ratio, collisionless merger remnants are characterized on average
by an anisotropy of $\delta \approx 0.35$ with a large spread. 
This value is close to the anisotropy found for the most flattened,
fast rotating SAURON ellipticals. The disagreement increases however for
less flattened, slow rotators. On average, even the fast rotating 
SAURON ellipticals have substantially
lower anisotropies and lower ellipticities than our
2:1 to 4:1 merger remnants. 

The disagreement is largest for slowly rotating systems.
Equal-mass (1:1) collisionless merger remnants, despite their slow rotation,
cannot reproduce at all the observed properties of the
slow rotators of the SAURON sample which are much less anisotropic and therefore
much less flattened than predicted by the numerical simulations.

One possible explanation for the disagreement between observations and
the theoretical models could be that the SAURON collaboration focussed
especially on axisymmetric systems, characterized by $\Pi_{yy}/\Pi_{xx} \approx 1$. 
We tested whether this could generate a bias towards preferentially spherical, low
anisotropy systems
by investigating the location of our axisymmetric merger remnants with
$\Pi_{yy}/\Pi_{xx} \geq 0.9$ in Fig.~2. Their distribution turns out not to be
different compared with the complete sample, ruling out such a solution.

\citet{2006MNRAS.372..839N} emphasised the importance of gaseous energy dissipation
during galaxy mergers.
They studied a set of disk mergers with mass ratios 1:1 and 3:1,
including gas with a mass fraction of 10 per cent of the total disk mass. The
gas dynamics was followed during the merging process, adopting an isothermal equation
of state. Star formation and stellar feedback was neglected. Despite this
simplification, \citet{2006MNRAS.372..839N} showed that a dissipative gas component,
settling into the galactic center through
its gravitational force has a strong effect on the orbital structure
of the merger remnant, leading to asymmetries of the
line-of-sight velocity distribution of rotating ellipticals that are in much
better agreement with observations than collisionless merger remnants (see also
\citealp{2006MNRAS.372L..78G}). The anisotropy
and ellipticity of the 1:1 and 3:1 merger remnants with 10\% gas however turns out
to be very similar to the distribution of the collisionless merger sample.
Clearly, a 10\% gas fraction as expected in evolved disk galaxies does not solve the problem
either.

Ellipticals are in general old systems that formed at a time when
disk galaxies were still quite gas-rich. Star formation and stellar as well
as central black hole feedback therefore is expected to have played an important role
during galaxy mergers \citep{2007arXiv0706.1246H,2007arXiv0706.1243H}. 
In order to investigate this question we have
started a new series of gas-rich ($\geq 20 \%$ gas) disk mergers, using GADGET2 and
taking into account star formation as well as stellar and black hole feedback
as described by \citet{2003MNRAS.339..289S} and \citet{2005ApJ...620L..79S}. 
A detailed analysis of these
simulations will be presented in a subsequent paper (Johansson et al., in preparation).
The triangles in Fig. 3 show how star formation and energetic
feedback affects the anisotropy and ellipticity of the merger remnants.
Large triangles correspond to simulations, including black hole accretion,
merging and black hole feedback. Five simulations have been repeated 
without taking into account black holes. They are represented in Fig.~3 by
the smaller triangles. For a more detailed investigation of how 
star formation, energetic feedback and black hole physics affects
the final structure of the merger remnants, table 1 compares
the anisotropies and ellipticities of the collisionless merger
simulations (columns 2 and 3) with those, starting with the same
initial geometries and mass ratios, however
now including 20\% of gas, star formation as well as stellar 
and black hole feedback (columns 4 and 5). The columns 6 and 7
finally show the results for the simulations where
black hole accretion and feedback has been neglected.
The initial conditions are shown in the first column of table 1 and are
defined in table 1 of \citet{2003ApJ...597..893N}.

\begin{table}
\begin{center}
\caption{Anisotropy parameter and ellipticity of merger remnants \label{tab:num}}
\begin{tabular}{l|c|c|c|c|c|c}
mass ratio/ &  $\epsilon_{int}$ & $\delta$    & $\epsilon_{int}$&  $\delta$        &$\epsilon_{int}$& $\delta$ \\
geometry  & stellar     &  stellar    & SF+BH&SF+BH& SF    & SF    \\
\hline
 1:1/7   & 0.49  &  0.41  &  0.36  &  0.20  &  &  \\
1:1/10  & 0.34  &  0.24  &  0.29  &  0.16  & 0.21 & 0.12 \\
1:1/13  & 0.44  &  0.38  &  0.27  &  0.24  & 0.28 & 0.24 \\
2:1/4   & 0.64  &  0.61  &  0.60  &  0.46  &  &  \\
2:1/10  & 0.24  &  0.07  &  0.47  &  0.37  &  &  \\
2:1/14  & 0.52  &  0.61  &  0.50  &  0.31  &  &  \\
3:1/4   & 0.69  &  0.47  &  0.63  &  0.44  & 0.56 & 0.39  \\
3:1/10  & 0.51  &  0.14  &  0.55  &  0.33  & 0.55 & 0.35  \\
3:1/14  & 0.59  &  0.42  &  0.61  &  0.24  & 0.62 & 0.25  \\
\hline
\end{tabular}
\end{center}
\end{table}

\vspace{0.3cm}

Fig.~3 and table 1 show, that star formation and stellar energetic feedback has
a strong effect on the anisotropy and ellipticity of merger remnants.
We still find a trend of decreasing rotational support, i.e. decreasing
$V/\sigma$ with decreasing mass ratio of the progenitor disks. In addition, now,
the scatter in the $\delta$ versus $\epsilon_{int}$ diagram
is much smaller and a clear correlation between $\delta$ and $\epsilon_{int}$
is visible that can be fitted by a linear relationship (dotted line)

\begin{equation}
\delta = 0.67 \epsilon_{int},
\end{equation}

\noindent which is somewhat steeper than the SAURON relation (solid line). 

It is interesting that including star formation, the location of the merger
remnants in the $\delta$-$\epsilon_{int}$ diagram shifts closer
to the dotted line, independent of whether they were above this
correlation or below it in the collisionless merger case.
For example, while
the 3:1 merger with initial geometry 10 in the collisionless 
case formed a remnant that is characterized by $(\epsilon_{int} / \delta) = (0.51/0.14)$, 
including star formation and black hole physics moves this system up to values of
$(\epsilon_{int} / \delta) = (0.55/0.33)$. Neglecting black holes, the values
are very similar with $(\epsilon_{int} / \delta) = (0.55/0.35)$.
An other example is the 1:1 merger with geometry 13 that in the collisionless
case is located at $(\epsilon_{int} / \delta) = (0.44/0.38)$ and that
with star formation shifts down to
$(\epsilon_{int} / \delta) = (0.27/0.24)$, much closer to the dotted line
than previously. In general, star formation is the dominant process.
As shown by the small triangles in Fig. 3 and table 1,
the effect of black hole accretion and feedback on the structure of
the merger remnants is small.

\begin{figure}[!ht]
\plotone{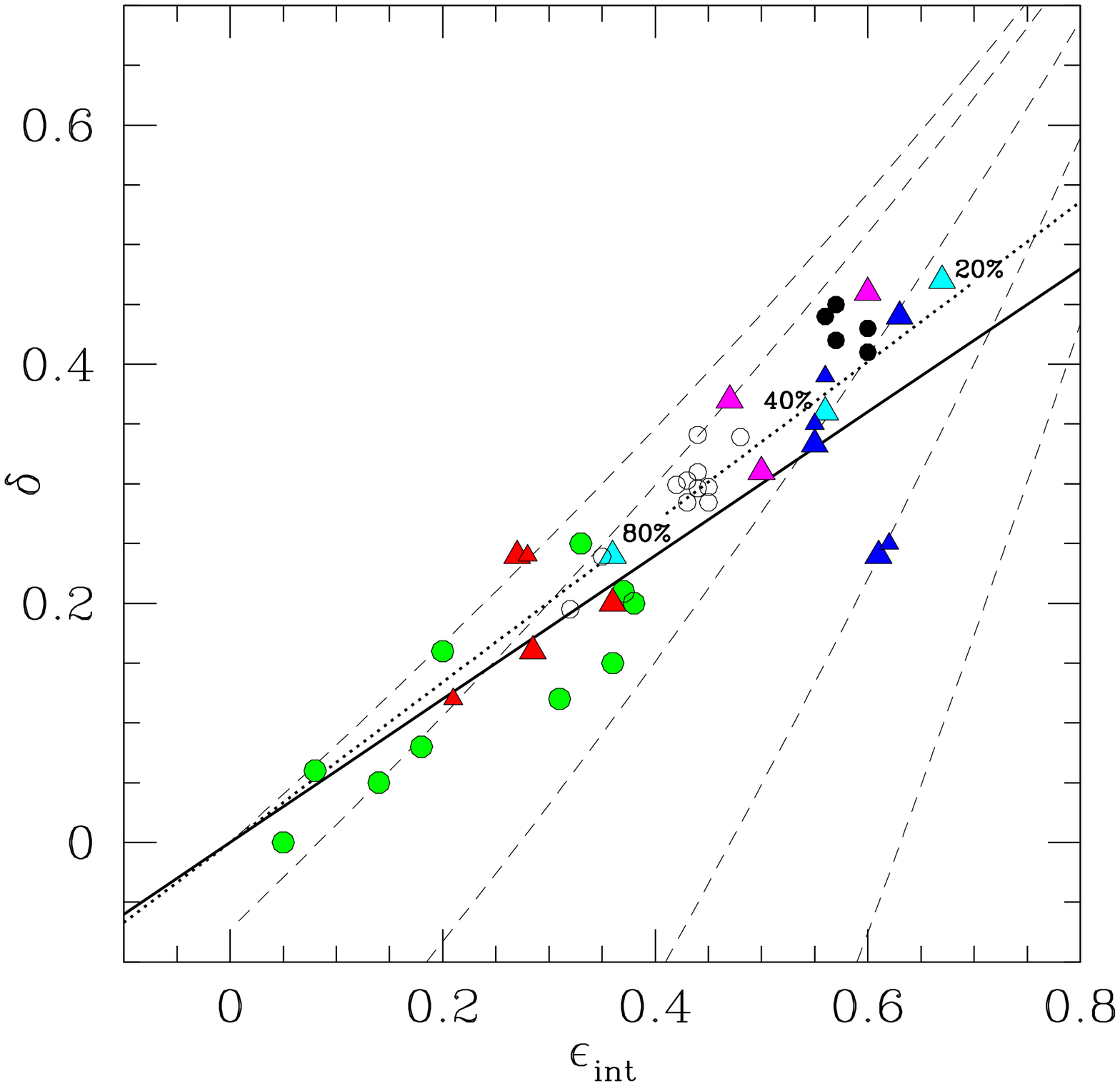}
\caption{
 \label{fig}
Same as Fig.~2. Here, red and blue large triangles show the location of 1:1 and 3:1
merger remnants with 20\% gas, including star formation and black hole accretion
as well as energetic feedback from stars and black holes.
The smaller triangles show merger simulations without black hole physics
(see also table 1).  Cyan triangles show the results of a co-planar disk merger, 
including star formation, black hole
physics and energetic feedback processes with initially 20\%, 40\% and
80\% gas, respectively. All data points can be fitted well by
a linear relationship of $\delta$ with $\epsilon_{int}$ that
is shown by the dotted line. It is somewhat steeper than the 
SAURON relation (solid line). Green circles correspond to cosmological
simulations of spheroidal galaxy formation that take into account star formation
but neglect stellar feedback and black hole physics. Filled black circles show
five dry equal-mass mergers of merger remnants that were produced from GADGET
merger simulations of disks with 20\% gas, taking into account star formation
and black hole physics. Open circles show the results of 1:1 and 3:1 dry mergers
of ellipticals that were generated by collisionless mergers of stellar disks
as discussed in \citet{2006ApJ...636L..81N}.}
\end{figure}

The influence of the initial gas ratio on the results is indicated by the cyan
triangles which show the results of three 3:1 mergers of two co-planar disk
galaxies with initial gas baryon fractions of 20\%, 40\% and 80\%. With increasing gas
fraction the remnants move along the dotted curve 
towards smaller ellipticities and anisotropies.

\section{The Origin of Slowly Rotating, Isotropic, Massive SAURON Ellipticals}

The merger simulations of disk galaxies with star formation still
cannot explain the almost round and
isotropic SAURON ellipticals with $\epsilon \leq 0.2$ and $\delta \leq 0.15$.
Mergers between ellipticals (dry mergers) have been suggested to dominate these
slowly rotating, red galaxies \citep{2003ApJ...597L.117K,2005astro.ph..6044F,
2006ApJ...636L..81N}.
A strong motivation for the dry merger scenario is the fact that old, red ellipticals
have masses that are much larger than typical spiral galaxies which basically rules out the possibility
that they could have formed by a binary spiral-spiral merger \citep{2007astro.ph..2535N}.

We investigated the dry merger scenario by re-merging the Gadget disk-disk
merger remnants, discussed in the previous paragraph. The results
of five 1:1 early-type mergers are shown by the filled black points in Fig.~3.
Although the explored parameter space is still very small it is already obvious
from the location of the black points that early-type mergers in general are unlikely
to explain the slowly rotating, isotropic SAURON ellipticals. The data points cluster
around $\epsilon_{int} \approx 0.57$ and $\delta \approx 0.43$ which is
close to the relationship between anisotropy and ellipticity (dotted line) found for
disk-disk mergers with star formation. From a dynamical point of view,
these dry merger remnants are similar to gas-rich disk mergers with star formation.
\citet{2006ApJ...636L..81N} had analysed the structure of ellipticals,
formed through dry re-merging of ellipticals that were generated from collisionless, 
1:1 and 3:1 stellar disk mergers as discussed in section 3.
Note that, as shown in Fig.~2, the progenitor ellipticals do not lie on the
SAURON relation. Nevertheless, it is interesting that re-merging of these systems
places them nicely on the $\delta$-$\epsilon_{int}$ relation described by
Eq.~7 (open circles in Fig.~3). Still, these merger remnants
are faster rotating, more anisotropic and more ellipsoidal than the slowly rotating
SAURON ellipticals.

Yet another possibility to generate spheroidal ellipticals are multiple
mergers in cosmological high-density regions. \citet{2007ApJ...658..710N} 
investigated the formation of a number of massive galaxies using
high-resolution cosmological simulations in a $\Lambda$CDM universe. The calculations
were simple, including only photoionization and cooling of the interstellar medium as well
as star formation. AGN and supernova feedback was neglected.
In these simulations efficient cooling
of gas generated rapid gas infall into smaller dark halo density perturbations,
followed by a burst of star formation. At the same time, these substructures
merged into massive spheroidal stellar galaxy, resembling a present-day, red
giant elliptical. Shock heating in the later phases generated a
surrounding hot gaseous halos that suppressed late star formation, leading
at the end to a red, old galaxy \citep{2003MNRAS.345..349B,2007MNRAS.380..339B}.

The green points in Fig.~3 show the location of three
spheroidal galaxies presented in \citet{2007ApJ...658..710N} and 7 additional
galaxies  of similar mass simulated in the same manner.
The distribution is in excellent agreement with the SAURON observations of red, massive
ellipticals. Interestingly, the simulations reproduce not only the observed low ellipticities
and anisotropies. They also show the same trend of anisotropy with ellipticity
as observed.

\section{Summary and Discussion}

The numerical simulations, discussed in the previous sections, have shown that interstellar
gas dynamics, star formation and stellar feedback plays a crucial role in order
to reproduce the observed kinematical and isophotal properties of fast rotating,
early-type galaxies. The final structure of the merger remnants
depends on the initial mass ratio
and gas fraction. The remnants are more round, less anisotropic and more
rotationally supported the smaller the mass ratio $M_1/M_2 \geq 1$ of 
the progenitors and the larger the initial gas fraction. The dependence 
of $\delta$ on $\epsilon_{int}$ is
in agreement with the observed trend found in the SAURON sample.

Subsequent dry re-merging of disk-disk 
merger remnants however does not generate the observed 
slowly-rotating red SAURON ellipticals
with small anisotropies and ellipticities. This indicates that at least some
early-type galaxies on the red galaxy sequence formed in a different way. We find
that multiple mergers of stellar substructures that formed
from cold gas infall into dark matter halos in cosmological simulations
produce round, isotropic and slowly-rotating relaxed stellar systems 
that are in perfect agreement
with the SAURON observations. Multiple mergers of
stellar systems in dense group enviroments therefore appear to be a promising
alternative scenario for the origin of the red, massive galaxy population.

Despite the fact that merger simulations with star formation
lead to a correlation between
anisotropy and ellipticity ($\delta = 0.67 \times \epsilon_{int}$)
that is very similar to that inferred from observations 
its origin is not understood yet. 
It is interesting that merger remnants appear to move
closer towards this relation along lines of constant $V/\sigma$
(i.e. roughly constant specific angular momentum) in the case of
a strong relaxation process. Here, strong relaxation is defined
as the merger of a system of kinematically cold systems of stars that lateron
break up and generate a kinematically hot stellar remnant. Several conditions
could lead to such a violent dynamical process. The cold stellar clumps could
for example have formed in the star-bursting tidal tails of interacting,
gas-rich disk galaxies. Another possibility is the cosmological multiple merging
of dark matter substructures with an embedded stellar systems.
The SAURON relation might represent the relaxed and phase-mixed end state of these
complex mergers with the location of the remnant on the relation being determined
by its specific angular momentum which is related to its value of $V/\sigma$.
More theoretical work will be required in order to better
understand these interesting questions and their connection to early-type galaxy
formation.

\begin{acknowledgements}
We thank M. Cappellari, R. Jesseit and J.P. Ostriker and Eric Emsellem for interesting 
discussions. The work was partly supported by the DFG Sonderforschungsbereich
375 "Astro-Teilchenphysik". The numerical simulations were run on a local
SGI-Altix 3700 Bx2 which was partly funded by the cluster of excellence
"Origin and Structure of the Universe".
\end{acknowledgements}

\bibliographystyle{apj}
\bibliography{./references}

\end{document}